\documentclass[a4paper,floatfix,rmp,twocolumn,showkeys,superscriptaddress]{revtex4}
  \usepackage[latin1]{inputenc}
  \usepackage{graphicx}
  \usepackage{mathtools}
  \usepackage{svg}
  
  \setcitestyle{numbers,square,comma}

\begin{document}

  \title{A spectrum of complexity uncovers Dunbar's number and other leaps in social structure}

  \providecommand{\NLin}{Group of Nonlinear Physics. Universidade de Santiago de Compostela, 15782, Santiago de Compostela, Spain. }
  \providecommand{\citmaga}{Galician Center for Mathematical Research and Technology (CITMAga), 15782 Santiago de Compostela, Spain}
  \providecommand{\FAp}{Departamento de F\'isica Aplicada and iMATUS, Universidade de Santiago de Compostela, 15782, Santiago de Compostela, Spain. }
  \providecommand{\CNB}{Systems Biology Department, Spanish National Center for Biotechnology (CSIC), C/ Darwin 3, 28049 Madrid, Spain. }
  \providecommand{\GISC}{Grupo Interdisciplinar de Sistemas Complejos (GISC), Madrid, Spain. }
  
  \author{Mart\'in Saavedra}
    \affiliation{\NLin}
    \affiliation{\citmaga}
  
  \author{Jorge Mira}
    \affiliation{\FAp}
  
  \author{Alberto P Mu\~nuzuri}
    \affiliation{\NLin}
    \affiliation{\citmaga}
  
  \author{Lu\'is F Seoane}
    \affiliation{\CNB}
    \affiliation{\GISC}

  \vspace{0.4 cm}
  \begin{abstract}
    \vspace{0.2 cm}

    Social dynamics are shaped by each person's actions, as well as by collective trends that emerge when individuals are brought together. These latter kind of influences escape anyone's control. They are, instead, dominated by aggregate societal properties such as size, polarization, cohesion, or hierarchy. Such features add nuance and complexity to social structure, and might be present, or not, for societies of different sizes. How do societies become more complex? Are there specific scales at which they are reorganized into emergent entities? In this paper we introduce the {\em social complexity spectrum}, a methodological tool, inspired by theoretical considerations about dynamics on complex networks, that addresses these questions empirically. We use as a probe a sociolinguistic process that has unfolded over decades within the north-western Spanish region of Galicia, across populations of varied sizes. We estimate how societal complexity increases monotonously with population size; and how specific scales stand out, at which complexity would build up faster. These scales are noted as dips in our spectra, similarly to missing wavelengths in light spectroscopy. Also, `red-' and `blue-shifts' take place as the general population shifted from more rural to more urban settings. These shifts help us sharpen our observations. Besides specific results around social complexity build-up, our work introduces a powerful tool to be applied in further study cases. 

  \end{abstract}


\maketitle

  \section{Introduction}
    \label{sec:1}
    
    \begin{figure*}
      \centering
      \includegraphics[width=0.7\textwidth]{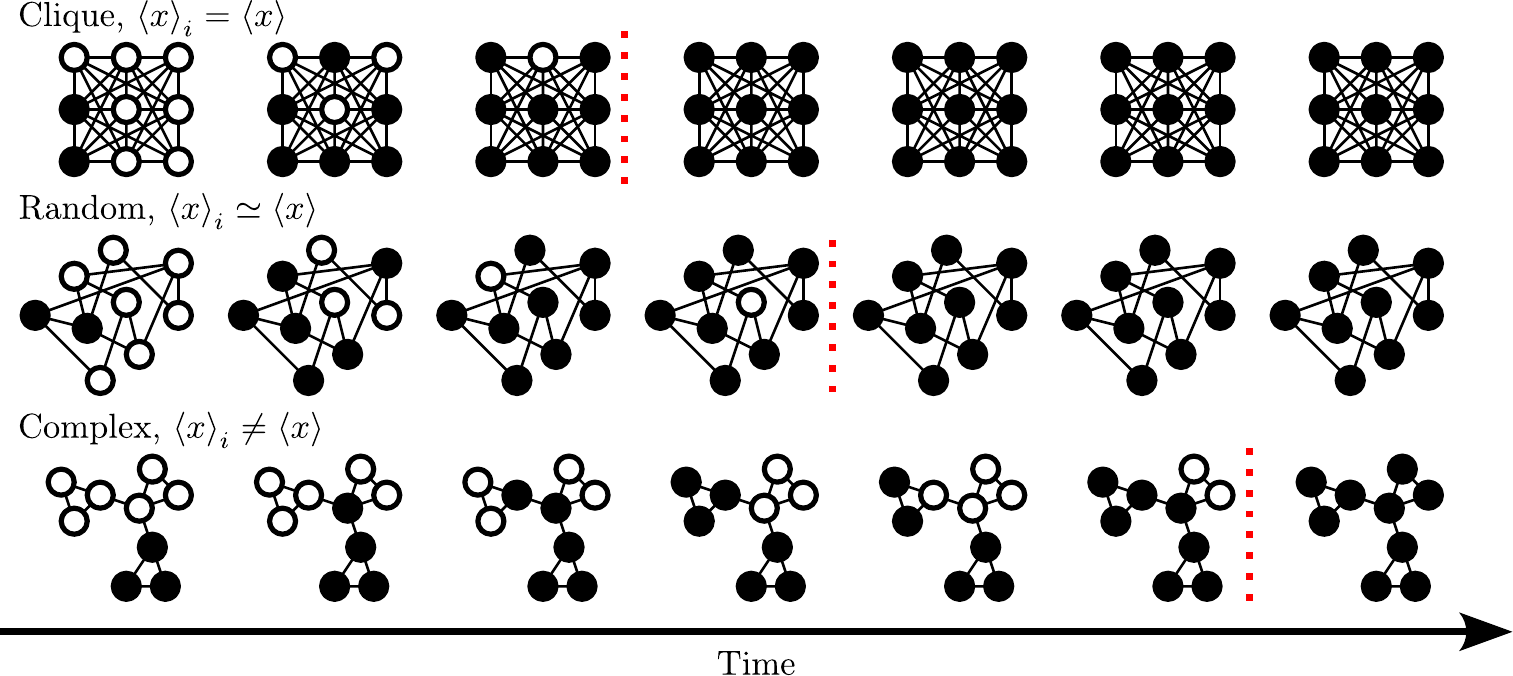}
      \caption{{\bf A complex network structure might slow down social dynamics.} In \cite{ToivonenSan2009} it is observed that certain opinion dynamics can take longer to reach their absorbing steady state due to a complex structure of the underlying social network. This figure illustrates a mechanism behind this phenomenon. In a clique (top), which is a simple kind of graph, all network individuals interact with each-other. This enables that the average opinion observed by the $i$-th individual equals the actual average opinion across the whole group, $\left<x\right>_i = \left<x\right>$. Conflicting views are resolved as quickly as possible (a red dashed vertical line marks when the steady state is reached in each case). In random, yet unstructured networks (middle), which are fairly simple as well, not everybody is connected with everybody else. However, a node's neighbors are relatively well distributed across the network, thus its perceived average opinion comes from a good-enough sampling of the group's state, $\left<x\right>_i \simeq \left<x\right>$. Dynamics are resolved only slightly more slowly. In complex, structured networks, individuals cluster locally with few nodes acting as gate-keepers of larger groups (bottom). Individuals inside each group perceive averaged opinions that might depart from the network's consensus, $\left<x\right>_i \ne \left<x\right>$. Hence, opinion dynamics are slowed-down as bottlenecks prevent the immediate invasion of each cluster. }
      \label{fig:0}
    \end{figure*} 

    If we put together grains of sand, one by one, at some point they become a mountain. When does that happen? Daniel Dennett reminds us that such questions might not have a clear-cut answer, and teaches us how to live with the ensuing scale of greys \cite{Dennett2013}. Yet some complex systems often change radically as their size crosses certain thresholds. At those points, emerging behaviors start shaping the fate of the whole, overriding the more straightforward dynamics of the parts. Then, as put beautifully by William P. Anderson, ``More is Different'' \cite{Anderson1972}. 

    As humans, we contemplate the duality of our autonomy and of belonging to an eusocial species. Tensions between both levels of Darwinian selection (the organismal and the societal) largely underlie our conflicting nature \cite{Wilson2014}. Individually, every modern human has likely experienced `being dragged by the masses'. But, when does a gathering of people become an emergent entity? How many such transitions might take place as societies grow? Are these processes gradual, as with mountains and grains of sand? Or can we spot relevant scales at which the dynamics of human behavior is altered by emerging levels of complexity? 

    We tackle these questions empirically through a process of opinion dynamics that has played out over social groups of different complexity. Specifically, we tend to socioliguistic processes of language shift that have occurred in the Spanish Autonomous region of Galicia over the last ${\sim}100$ years \cite{MussaMira2019}. In this process, the vernacular, Galician, trended to be substituted by Castilian Spanish. Both are romance languages with high mutual intelligibility, which might enable long-term bilingualism and coexistence of both tongues \cite{MiraNieto2011}. These dynamics have unfolded simultaneously, at different speeds, in a range of population centers, from very rural ones to larger cities. Galicia presents quite unusual demographics: It covers about $6\%$ of the territory of Spain and houses a similar percentage of the country's population. Notwithstanding, Galicia contains roughly a half of all $60\>000$ Spanish Singular Population Entities (SPEs, defined as any unit such as villages, cities, etc.). Among these, as of $2016$, 27 000 Galician SPEs had less than $100$ inhabitants \cite{INE}. This gives us a great sampling of social processes happening on communities with different structures. If we assume that SPEs of different sizes constitute underlying social groups of varying complexity, we are provided with a unique opportunity to study at what emerging social scales human behavior is altered in a noticeable way. 

    Social exchanges happen within a social substrate, often modeled as a graph or network. In them, people are nodes, and edges represent existing interactions---whether virtual or of physical contact. Networks can have different overall structures---e.g., {\em small world} \cite{WattsStrogatz1998}, hierarchical or more horizontal \cite{CorominasMurtra2013}, etc. Distinct network structures enable or hinder the unfolding of different phenomena, such as opinion dynamics, epidemic spread, coordination towards a goal, etc. For example a sufficiently sparse network of physical contacts can halt the spread of a virus---as confinement measures during the recent pandemic have illustrated. Less trivially, Darwinian evolution can be accelerated if it happens over networks with specific shapes \cite{AdlamNowak2015, TkadlecNowak2021}. 

    A similar interplay between network structure and dynamics can apply to other social processes---specifically, to opinion dynamics such as the decision to keep or change your tongue. In a computational study, Toivonen et al.\ \cite{ToivonenSan2009} showed how certain classes of opinion dynamics have longer relaxation times in more complex networks, while the same processes converge faster to their steady states in simpler graphs. In more trivial networks (Fig.\ \ref{fig:0}, top and middle), each individual samples accurately the social group's average opinion, and dynamics can be resolved quickly. In more complex graphs (Fig.\ \ref{fig:0}, bottom) nodes of similar opinion can form clusters guarded by gatekeepers that hardly flip their opinions. These clusters become difficult to penetrate. Hence, some individuals are cut out of the emerging consensus, and convergence to a homogeneous opinion is hindered. Inspired by this, it was found empirically that sociolinguistic dynamics in Galicia have unfolded faster in more rural areas, and slower in more urban ones \cite{MussaMira2019}. 

    This last study assumed that urban communities had a more complex social structure, and it used the threshold of $5\>000$ inhabitants to consider a SPE as {\em urban}. This choice was based on Spanish legislation \cite{BOE} that demands that counties (which usually include several SPEs) over that size present a series of structures and services (a public park and library, a market place, and a waste management system), potentially marking a jump in social network complexity. Further demands of urban equipment do not happen below $5\>000$ inhabitants, or above until $20\>000$ and $50\>000$ inhabitants. However, can we relax this definition of {\em urban} and come up with a more organic way to find salient leaps in social complexity? Perhaps, if such changes in complexity affect ongoing opinion dynamics (as those simulated by Toivonen et al.\ \cite{ToivonenSan2009}), we can look at empirical data of language shift to spot the relevant scales at which complexity builds up. 

    In this paper we study this possibility. We try a range of scales (based on population sizes), and check whether each scale is a good separation between simple versus complex social networks. Following \cite{ToivonenSan2009}, our test is whether sociolinguistic dynamics tend to play out faster in `simpler' networks. We measure this through correlations between a region's purported complexity and the rate at which sociolinguistic dynamics have unfolded in that area. This renders a kind of `spectrogram' in which dips are visible at population scales with non-trivial correlations (much like a spectral line is missing in optical spectroscopy when light traverses specific chemical compounds that absorb a specific wavelength). We find two salient scales that, according to our criterion, would separate simpler from more complex social networks. Our results suggest that at those scales some emergent component takes over and alters the pace at which sociolinguistic dynamics play out. One of these scales corresponds to the threshold used in \cite{MussaMira2019}, which was based on urban planning. The other one, more prominent, happens at much smaller community sizes (${\sim}200$ people). It does not correlate with any feature marked by Spanish law. Rather, its proximity to Dunbar's number (an empirical---yet debated \cite{deRuiter2011, Wellman2012, Lindenfors2021}---cognitive limit to the number of relationships that animals can have \cite{Dunbar1992, Dunbar1998, Dunbar2015, Carron2016, TamaritSanchez2018, Dunbar2020a, TamaritCuesta2022, BzdokDunbar2022}) suggests an organic emergence of social complexity. 

    In Sec.\ \ref{sec:meth.1} we describe the sociolinguistic dynamics of interest to us and the empirical data available. In Sec.\ \ref{sec:meth.2} we introduce the mathematical model that we fit to the data. The fitting procedure is explained in Sec.\ \ref{sec:meth.3}. We use the resulting model parameters to estimate the rate at which the dynamics unfold in each geographical region. In Sec.\ \ref{sec:res.1} we introduce our {\em social complexity spectrum}. We explain how we define separations between potentially simple and potentially complex social networks, and how we evaluate the goodness of these separations. Secs.\ \ref{sec:res.1} and \ref{sec:res.2} contain our main results. Namely, that we identify two outstanding scales at which leaps, or buildups of social complexity would happen according to our criterion. Sec.\ \ref{sec:res.3} further explores the {\em spectrum of social complexity} as a methodological tool. Similarly to physics and optical spectroscopy, we observe `red-' and `blue-shifts' as the overall demographics has changed over two decades. We illustrate how this can help us refine the separation of simple and complex social networks. We wrap up the paper with a discussion of our findings in Sec.\ \ref{sec:disc}. We further argue that the ``complexity spectrum'' might be a powerful tool to uncover scales of social relevance when applied to similar dynamics. 

    \begin{figure*}
      \centering
      \includegraphics[width=\textwidth]{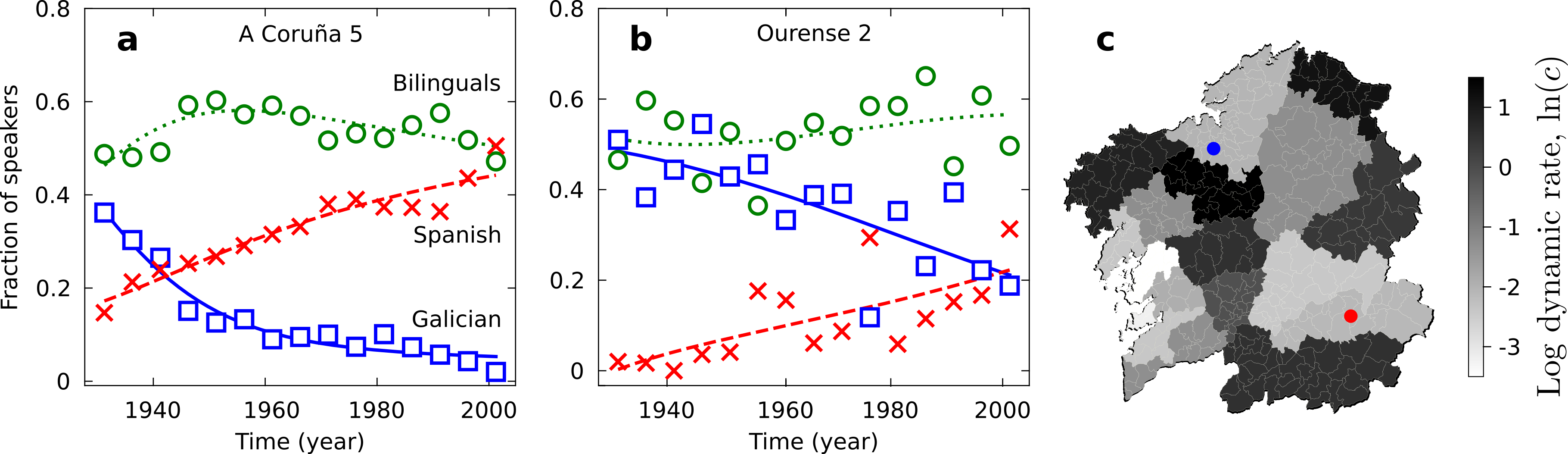}
      \caption{{\bf Fitting model to empirical data.} {\bf a} Empirical data and model fit for region {\em A Coru\~na 5}, which results in the best-quality fit (smallest $\chi^2$ value). {\bf b} Same for region {\em Ourense 2}, which results in the worst-quality fit. {\bf c} Map of all Galician regions colored after the logarithm of the dynamic rate, $\ln(c_i)$. Best -and worst- quality fit regions are marked respectively with blue and red dots. }
      \label{fig:1}
    \end{figure*} 

  \section{Methods}
    \label{sec:meth}

    \subsection{Sociolinguistic dynamics and data}
      \label{sec:meth.1} 

      We investigate the emergence of singular scales of social complexity by looking at how certain sociolinguistic dynamics have unfolded at different speeds in regions that, potentially, contain distinctly complex social networks. The dynamics that we use as a proxy is the coexistence of Castillian Spanish and Galician. Both these tongues are romance languages that coexist in the Autonomous Community of Galicia, in north-western Spain. Mutual intelligibility between them is large, allowing broad bilingual communities and, potentially, a sustained coexistence between the two of them. While Galician is the vernacular, a shift towards Castillian Spanish has been underway for centuries, and has been especially accentuated during the 20th century. 

      The Galician Statistical Office ({\em Instituto Galego de Estat\'istica}, IGE) has tracked language use in different Galician regions and across demographic groups. In periodic polls, informants would self-assess their language use as `only Galician', `mostly Galician', `same use of both tongues', `mostly Spanish', and `only Spanish'. We took informants at either extreme of this scale as monolingual individuals of the corresponding language, and grouped the central categories as bilinguals. We are interested in the fractions of speakers in these groups. 

      IGE polls are stratified by age, which allows us to build a time-series of fractions of speakers by projecting age groups in apparent time \cite{Labov1963, Chambers2013, Eckert1997, Mague2006, SeoaneMira2019} (meaning that the fraction of speakers of people of a certain age become estimators of the fraction of speakers when those people were born). Additionally, data is split into $20$ independent Galician subregions, each of which is made up of a collection of smaller counties---but data for individual counties is not available. Hence, our empirical dataset of sociolinguistic dynamics consists of $20$ time series with the fractions of monolingual Galician speakers, bilingual speakers, and monolingual Spanish speakers (Fig.\ \ref{fig:1}{\bf a-b}). We base our work on the IGE poll that allowed us of to estimate the fractions of those born in $2001$ \cite{ige2001}.

    \subsection{Mathematical model}
      \label{sec:meth.2}

      Beginning in the early $90$s \cite{BaggsFreedman1990, BaggsFreedman1993}, a growing community of mathematicians, physicists, ecologists, and complexity researchers started using systems of differential equations to model possible trajectories of speakers of coexisting languages over time \cite{SeoaneMira2017}. A turning point was the work by Abrams and Strogatz \cite{Abrams2003}, who fitted their equations to data from dozens of cohabiting tongues. This inspired a wave of new models whose stability and dynamical classes were thoroughly analyzed \cite{MiraParedes2005, PinascoRomanelli2006, CastelloSan2006, MinettWang2008, PatriarcaHeinsalu2009, MiraNieto2011, OteroMira2013, CastelloSan2013, HeinsaluLeonard2014, ColucciOtero2014, ColucciMira2016, LoufRamasco2021}, and which could in some occasions be tested against empirical data \cite{SeoaneMira2019, LoufRamasco2021, ZhangGong2013, KandlerSteele2008, Kandler2009, KandlerSteele2010, IsernFort2014, ProchazkaVogl2017, MussaMira2019, MimarGoshal2021, SeoaneMira2021}. 

      Different authors would emphasize distinct ingredients that might affect language coexistence, such as their spatial distribution, or bilingualism (elements that the original model by Abrams and Strogatz did not contemplate). We use one such variation that includes bilingualism \cite{MiraParedes2005}, whose stability and dynamics have been studied in detail \cite{MiraNieto2011, OteroMira2013, ColucciOtero2014, ColucciMira2016}, and that has been fitted to data of different cohabiting languages, including the Galician-Spanish case \cite{SeoaneMira2019, MussaMira2019, SeoaneMira2021}. Contrary to some other models with bilingualism, the one that we use is compatible with either the stable coexistence of both tongues, or that one language takes over and drives the other to extinction. Thus, the model is agnostic regarding the stability of the coexisting couple, and the empirical data can constrain model parameters towards either outcome. 
    
      The model consists of a system of coupled differential equations that tracks the time-evolution of the fraction, $x$, of monolinguals of language X (here, Galician); of the fraction, $b$, of bilinguals; and of the fraction, $y$, of monolinguals of language $Y$ (here, Spanish). Population is normalized such that $x + y + b = 1$, hence two equations suffice to solve the system. These equations read:
        \begin{eqnarray}
         {dx \over dt} &=& c\left[s\left(1-k\right)\left(1-x\right)\left(1-y\right)^a-x\left(1-s\right)\left(1-x\right)^a\right], \nonumber \\
         {dy \over dt} &=& c\left[\left(1-s\right)\left(1-k\right)\left(1-y\right)\left(1-x\right)^a-ys\left(1-y\right)^a\right]. \nonumber \\
         \label{eq:1}
        \end{eqnarray}
      This is a compact form of a simple normalized flow between monolinguals $X$ and $Y$ and the bilingual group $B$. Details regarding the interpretation of the equations and parameters can be found in the model's literature \cite{SeoaneMira2017, MiraParedes2005, MiraNieto2011, OteroMira2013, ColucciOtero2014, ColucciMira2016, SeoaneMira2019, MussaMira2019, SeoaneMira2021}. 

      However, in a nutshell: The likelihood that a speaker of a tongue (say $X$) starts using the other one (hence $Y$) is proportional to the {\em prestige}, $s_Y$, of the target language and to a monotonous function of the target fraction of speakers. The prestiges ($s_X$ and $s_Y$) represent coarse grained attractions exerted by either tongue. They are normalized ($s_X + s_Y=1$), such that it suffices to track $s \equiv s_X$. The monotonous function of the target fraction of speakers is usually an attractive term that makes the opposite language more appealing the more people that use it. Say, a monolingual speaker of $X$, by switching to $Y$ completely, can communicate with a population of size $1-x$. In the model, the appeal of that attractive population is modulated by an exponent $a$, such that the flux away from $X$ becomes proportional to $(1-x)^a$. Same reasoning applies for fluxes away from $Y$. The parameter $a$ has been termed {\em volatility} \cite{CastelloSan2013, SeoaneMira2017}, implying that it captures how stable established groups are. If $a > 0$, larger groups tend to exert bigger attractive forces. If $a=0$, the size of the target population when acquiring a new language does not matter. The case $a<0$ has been studied purely mathematically \cite{ColucciOtero2014}, and termed an `exotic' scenario since it implies that speakers of a majority language tend to abandon their own tongue. Finally, of all speakers acquiring the opposite language, a fraction $k$ retains both and a fraction $1-k$ becomes purely monolingual in the target tongue. Since the larger $k$ is, the easier it is to retain both languages, this parameter has been termed {\em interlinguistic similarity}. The parameter is extracted from fits to data, thus we prefer to interpret it as a coarse grained, emergent, or effective {\em interlinguistic similarity} that might factor in multiple circumstances (e.g.\ languages might be grammatically distant, but social and economic constraints might increase the appeal and ease to keep them both \cite{SeoaneMira2021}). 

      These three parameters, $(a, k, s)$, together with the initial conditions (i.e.\ the fractions of speakers in each group at an arbitrary time $t=0$), affect the qualitative trajectories of speakers over time. Depending on their specific values, one language might drive the other to extinction, or both might coexist asymptotically along with a bilingual group. Depending on the values of the model parameters, the qualitative outcome might or might not depend on the initial conditions. A final model parameter, $c$, does not affect the long-term stability as it just measures the speed at which the dynamics unfold. This is the most important parameter for us.

    \subsection{Fitting the mathematical model to empirical data}
      \label{sec:meth.3}

      Let us label the $N=20$ Galician regions for which we have independent data as $\{ R_i,\>i=1, \dots, 20\}$. Each region provides an empirical time series for the fractions of speakers in each group: $\bar{x}_i \equiv \{ \bar{x}_i(t),\>t = t^0, \dots, t^{end} \}$ and $\bar{y}_i \equiv \{ \bar{y}_i(t),\>t = t^0, \dots, t^{end} \}$. The fraction of bilingual speakers is ignored, as we get it trivially from the normalization of the population. Projection in apparent time yields $15$ samples from years $t^0=1931$ to $t^{end}=2001$, separated by $5$-year intervals. For each region, our task is to find the set of parameters, $\Pi_i \equiv \{c_i;\>a_i, k_i, s_i;\>x^0_i, y^0_i\}$, that result in the time evolution that more accurately tracks that of $\bar{x}_i(t)$ and $\bar{y}_i(t)$. Note that we include the initial conditions, $x^0_i$ and $y^0_i$, as parameters to be fitted. Alternatively, we could have taken $x^0_i \equiv \bar{x}_i(t^0)$ and $y^0_i \equiv \bar{y}_i(t^0)$; but we found that our procedure is robust and fast enough to handle initial conditions as well. 

      To find the best set of parameters for each region we seek to minimize the target function: 
        \begin{eqnarray}
           \chi_i^2 (\Pi_i) &=& \frac{1}{2} \sum_t \Big[ \big( \bar{x}_i(t) - x_i(t;\Pi_i) \big)^2 \nonumber \\ 
                            && + \big( \bar{y}_i(t) - y_i(t;\Pi_i) \big)^2\Big]. 
           \label{eq:2}
        \end{eqnarray}
      Note that $\bar{x}_i$ and $\bar{y}_i$ are empirical data while $x_i$ and $y_i$ correspond to variables of the model. Our numerical minimization of $\chi_i^2(\Pi_i)$ starts out by choosing a random seed of parameters with $c_i \in [0,1)$, $a_i \in [0, 2)$, $k_i \in [0, 1)$, and $s_i \in [0,1)$. Additionally, random initial conditions are generated with the constraints $x^0_i \in [0,1)$ and $y^0_i \in [0,1-x^0_i)$. We then approximate the gradient of $\chi_i^2$ around these values by evaluating the effect in $\chi_i^2$ of slight variations on each parameter: 
        \begin{eqnarray}
           \nabla_\pi \left(\chi_i^2\right) &=& {\chi_i^2(\pi+\Delta \pi)-\chi_i^2(\pi-\Delta \pi) \over 2 \Delta \pi}; 
           \label{eq:gradient}
        \end{eqnarray}
      where $\pi \in \Pi$ is one of our $4$ parameters or $2$ initial conditions, and $\Delta \pi = 10^{-4}$ implements the slight perturbation of each parameter. Each evaluation of $\chi_i^2$ (i.e.\ each $\chi_i^2(\pi+\Delta \pi)$ and $\chi_i^2(\pi-\Delta \pi)$ for each $\pi \in \Pi$) entails a $4^\text{th}$-order Runge-Kutta numerical integration of the system of Eqs.\ \ref{eq:1}. Eq.\ \ref{eq:gradient} is evaluated for each $\pi \in \Pi$ independently. The result of these evaluations is fed to the Adam algorithm \cite{KingmaBa2014}, which returns an improved estimation of the gradient $\nabla_\pi^{Adam} \left(\chi_i^2\right)$. We proceed to update each parameter as $\pi \rightarrow \pi- \eta \nabla_\pi^{Adam} \left(\chi_i^2\right)$, with $\eta = 0.0001$. We iterate this process until a satisfactory convergence (the algorithm halts when $\chi_i^2$ has not changed more than $10^{-7}$ in relative value within $100$ iterations). 

      For each region, we launched $1\>000$ parallel optimizations of $\chi_i^2$ with different initial seeds. Most minimization processes converged robustly towards the same (numerically indistinguishable) final values of the model parameters and initial conditions. Their average became our $\Pi_i \equiv \{c_i;\>a_i, k_i, s_i;\>x^0_i, y^0_i\}$ for each region. A few processes were halted before convergence because they were taking too many iterations---they were likely stuck around low-quality local minima. \\

      Realistic values of model parameters are constrained to: $0 \le k, s \le 1$ and $c > 0$. Other values of $k$ and $s$ can result in negative fractions of speakers, and negative values of $c$ would see the dynamics running backwards. The final parameter has usually been studied with $a>0$. This has been especially so when fitting the model to empirical data, since negative values of $a$ mean that speakers would be repelled by their own linguistic group if it is the largest one \cite{ColucciOtero2014}. All fits to data fall within these constraints except for $4$ regions (marked with asterisks in Sup.\ Fig. 1{\bf a}), in which we find negative values of $a$. In these cases, $a$ can be varied in a range that includes $a>0$ and (i) all other parameters remain relatively unchanged and (ii) the fit does not result in much worst values of $\chi^2$. 

      For the results in the main part of the paper, we constrained fits in those $4$ regions to present positive values of $a$ (which is more reasonable from a sociolinguistic viewpoint). Therefore, we fitted data in those regions allowing only updates in model parameters ($c_i$, $a_i$, $k_i$, and $s_i$), using $x^0_i = \bar{x}_i(t^0)$ and $y^0_i = \bar{y}_i(t^0)$ as initial conditions. This converged to values of $a>0$ in all cases. Then, separately, we refined the fit allowing updates on $x^0$ and $y^0$ only. In the Supporting Material we show all calculations with the parameters obtained allowing $a<0$. The overall results remain similar. We also report what happens if we remove the $4$ pathological regions. This had a non-trivial effect in our outcome. 

      After the fitting procedure, we are left with a reconstruction of the sociolinguistic dynamics over time (Fig.\ \ref{fig:1}{\bf a-b}) for each region, $R_i$. Each trajectory is given by a set of parameters $\Pi_i \equiv \{c_i; a_i, k_i, s_i; x^0_i, y^0_i\}$. Of these, we are interested in the values, $c_i$, that determine how fast or slow the dynamics unfold in each region (Fig.\ \ref{fig:1}{\bf c}). Note that changing this parameter does not alter whether both languages survive in the long term, or whether one drives the other to extinction. This parameter only speeds up or slows down the time evolution. In any case, here we are not concerned with the long-term fate of the linguistic cohabitation. 

      The fitting procedure used in this paper differs from that in \cite{MussaMira2019}. The current one converges to more optimal parameter values. This explains some numerical differences---which are expected from stochastic processes nevertheless. The results from \cite{MussaMira2019} are robust to these changes.

  \section{Results}    
    \label{sec:res}

    \begin{figure*}
      \centering
      \includegraphics[width=\textwidth]{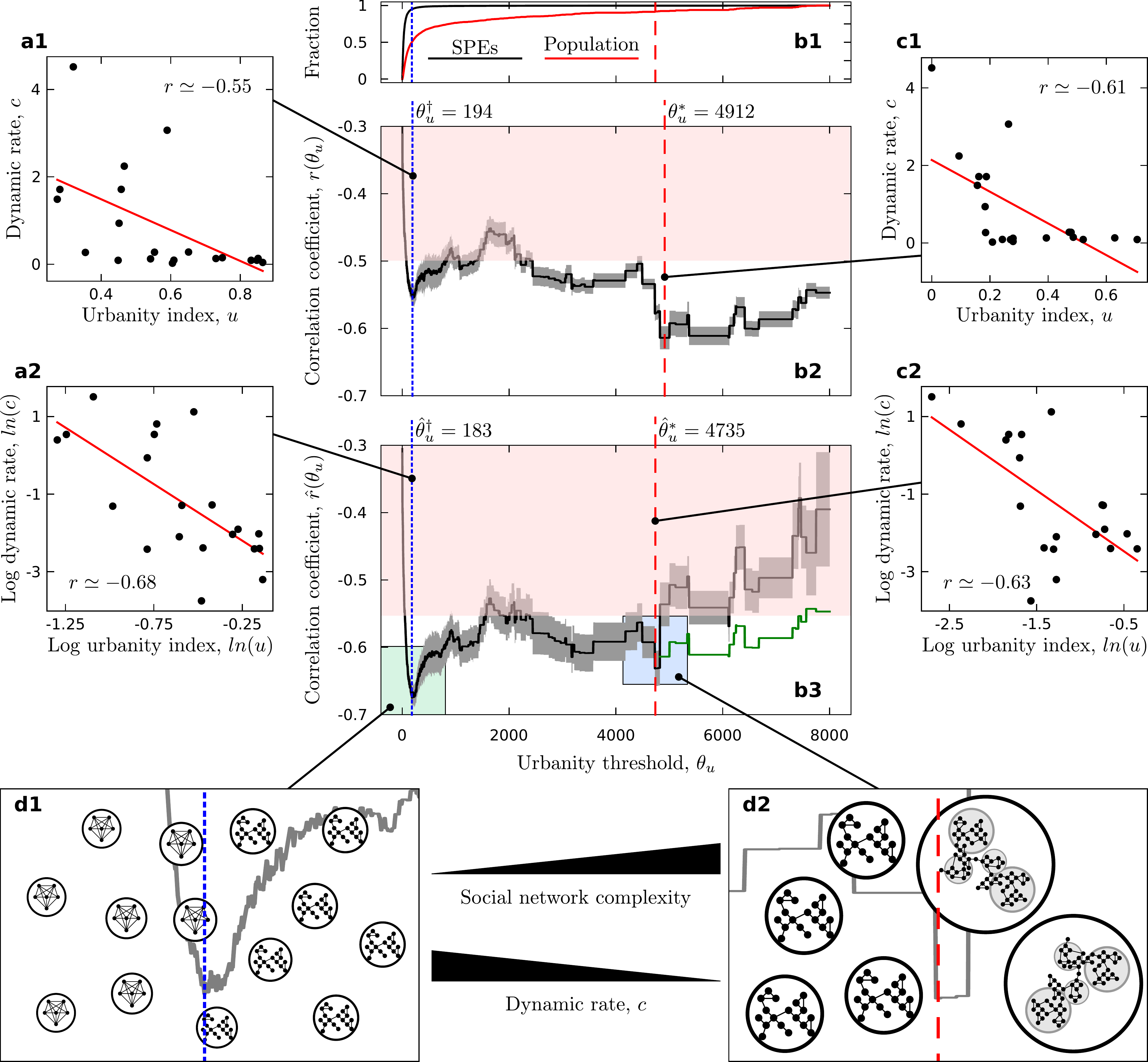}
      \caption{{\bf Social complexity spectra.} {\bf a} Examples of correlations between urban indexes and dynamic rates, taking $\theta_u$ as the local minima in the prominent dip at low population sizes. These local minima are marked by $\theta_u^\dag = 194$ for the linear spectrum ({\bf a1}) and $\hat{\theta}_u^\dag = 183$ for the logarithmic spectrum ({\bf a2}). {\bf b1} The black curve traces the fraction of SPEs with population below a given $\theta_u$. The red plot marks the fraction of Galician population living in SPEs sized less than $\theta_u$. This panel shows that nothing suspicious, which could trivially explain the properties of our spectra, happens at the outstanding scales (marked with vertical lines). {\bf b2} Social complexity spectrum (black curve) assuming the straightforward, linear relationship $c \propto u$. Standard deviation of the spectrum (gray shading) was estimated by $10$-fold jackknifing \cite{Jackknife}. A red dashed line marks $\theta_u^* = 4912$ and a blue dotted line indicates $\theta_u^\dag = 194$. A null model was implemented by reshuffling all SPEs to the $20$ regions. This resulted in a spectrum centered around $r(\theta_u) \sim 0$. The red-shaded area marks two times the worst-case standard deviations found in that model (see Sup.\ Fig.\ 2 for typical standard deviations, which are much more lenient). Thus, $r(\theta_u)$ values below the shading have a chance less than $0.023$ of happening by chance. {\bf b3} Same for the logarithmic spectrum, which assumes $c \propto u^\beta$. The red dashed line marks $\hat{\theta}_u^* = 4735$ and the blue dotted line indicates $\hat{\theta}_u^\dag = 183$. Areas shaded in green and blue are amplified in {\bf d1} and {\bf d2} respectively. {\bf c} Same as in {\bf a}, but for the spectral dip at large population size, with local minima marked at $\theta_u^* = 4912$ ({\bf c1}) and $\hat{\theta}_u^* = 4735$ ({\bf c2}). {\bf d} Cartoons illustrating the main implications of our hypotheses: A scale at which social complexity builds up very quickly segregates simpler from more complex social networks. Dynamics being slower in more complex networks would result in a good, negative correlation between dynamics rates within a region and that region's {\em urbanity}. We portray two distinct jumps in complexity to suggest that the social complexity buildup at around $\hat{\theta}_u^\dag$ ({\bf d1}) is likely very different from the one around $\theta_u^*$ ({\bf d2}). }
      \label{fig:2}
    \end{figure*} 

    \subsection{Building a spectrogram of social complexity}
      \label{sec:res.1}

      In \cite{ToivonenSan2009}, Toivonen et al.\ found, in simulations, that certain (not all) kinds of opinion dynamics relax faster to their steady state when they unfold in more complex graphs. Specifically, they simulated opinion dynamics on (i) rather simple random networks, and on (ii) more complex graphs that have a mesoscale because they consist of randomly connected cliques. In the former, the dynamics tend faster to their absorbing state. In the later, the lifetimes of transitory states of certain ordering dynamics present broad distributions---meaning that they might take arbitrarily long times to settle into a final, stable configuration. The kind of ordering dynamics studied in \cite{ToivonenSan2009} consist of nodes that change their state depending on the preferences of their neighbors---much like speakers might change their tongue influenced by others whom they communicate with. Indeed, the dynamics in \cite{ToivonenSan2009} have been used to simulate language shift in networks of connected speakers. 

      Inspired by this, Mussa Juane et al.\ \cite{MussaMira2019} reasoned that sociolinguistic dynamics might unfold slower in more complex interaction networks. They also hypothesized that the amount of urban-dwelling inhabitants of a region might capture the complexity of the networks that would underlie social interactions in the area. If all these assumptions were correct, the rate at which sociolinguistic dynamics proceed should correlate negatively with the fraction of urban dwellers---as they empirically found. 

      The speed of sociolinguistic dynamics in \cite{MussaMira2019} was measured similarly to how we do it here---by fitting the model of language shift from Eqs.\ \ref{eq:1} to our same dataset, and considering the parameter $c_i$ for each Galician region. The only effect of this parameter on the dynamics is to tune their velocity. 

      For the fraction of urban dwellers, Mussa Juane et al.\ considered an urbanity threshold, $\theta_u$, and counted the fraction of people in each region that lived in SPEs above $\theta_u$. To better formalize this, consider all Galician SPEs, and let us label them $\{\nu_j,\>j=1, \dots, N^{SPE}\}$, with $N^{SPE}$ the total number of Singular Population Entities in Galicia. Let us also label SPEs within a single region as $\{\nu_i^k,\>k=1, \dots, N^{SPE}_i\}$, with $N^{SPE}_i$ the number of SPEs in region $R_i$. We then define the urbanity index of this region as: 
        \begin{eqnarray} 
          u_i &=& {\sum_k H(\nu_i^k)\cdot\Theta\left(H(\nu_i^k) - \theta_u\right) \over \sum_k H(\nu_i^k)}; 
          \label{eq:defUrban} 
        \end{eqnarray}
      where $H(\nu_ i^k)$ returns the number of inhabitants in $\nu_i^k$, and $\Theta(\cdot)$ is Heaviside's theta function (which, in this case, is $1$ if $\nu_i^k$ has $\theta_u$ or more inhabitants, and $0$ otherwise). Thus, $u_i$ is the fraction of people in region $R_i$ that live on SPEs with more than $\theta_u$ inhabitants. Information about the number of inhabitants in each SPE is provided yearly by the IGE since $1999$ \cite{ige2022}. For the remainder of this section and in section \ref{sec:res.2}, we use the data from $2002$. This choice is justified, and then relaxed, in section \ref{sec:res.3}. 

      Mussa Juane et al.\ took a value $\theta_u \equiv 5\>000$ guided by Spanish law, which states that counties (which, however, might include several SPEs) above this size must offer a series of services---specifically: a public park and library, a market place, and a waste management system. The hope was that this threshold would separate between simpler and more complex SPEs, and hence that a negative correlation between $c_i$ and $u_i$ should be observed---as it was the case. \\ 

      We move beyond this and wonder whether other scales, different from $\theta_u \equiv 5\>000$, might segregate simpler from more complex social networks as well. Therefore, we make the urbanity index an explicit function of the threshold, $u_i \equiv u_i(\theta_u)$, and explore how the correlation $r(\theta_u)$ between $c_i$ and $u_i(\theta_u)$ changes as a function of our varying definition of urban, $\theta_u$. Figure \ref{fig:2}{\bf b2} plots this behavior. We term such representation a {\em spectrogram of social complexity} or a {\em complexity spectrogram}. 

      We might expect that more populated SPEs are naturally more complex than SPEs with less inhabitants. In absence of other processes, the sheer combinatorial opportunity of more interactions might suffice. We do not discard that large enough cities might drain the complexity of its constituent parts, resulting in simpler networks. We come back to this possibility in the Discussion. But let us put this case aside for a moment and consider that complexity tends to increase rather monotonously with population size. Then, if the main hypotheses in this paper hold true, we would expect $r(\theta_u) < 0$ for any $\theta_u$, as any scale would separate, in average, simpler from more complex social networks. Fig.\ \ref{fig:2}{\bf b2} shows that this is the case. This result is replicated in all our complexity spectra. We do not expect a perfect correlation, given all the sources of noise in the process---e.g.\ urban areas might have declined yet remain complex, the sociolinguistic dynamics is inherently stochastic, etc. 

      A second possibility is that, at a certain scale, $\theta_u^*$, social complexity builds up rather suddenly. If that is the case, SPEs with populations below such size, $H(\nu_j) < \theta_u^*$, should be notably simpler than SPEs above the threshold (as illustrated in Figs. \ref{fig:2}{\bf d1} and {\bf d2}). In such case, $\theta_u^*$ should mark a rather good separation between simpler and more complex social networks, and correlation between the speed of social dynamics, $c_i$, and the urbanity index at that scale, $u_i(\theta_u^*)$, should correlate notably better than for other values of $\theta_u \ne \theta_u^*$. This should be noted as a dip (accompanied by a local or global minimum) in the complexity spectrum---reminding us of gaps in optical spectroscopy after light of a specific wavelength has been absorbed. 

      The global minimum of the spectrum in Fig.\ \ref{fig:2}{\bf b2} is found at $\theta_u = \theta_u^* \equiv 4\>912$ inhabitants. The complexity spectrum indeed sees a steep descent as this threshold is approached from the left. This would indicate, within our framework, that choosing a threshold below $\theta_u^*$ would be a much worst separator between simpler and more complex SPEs. However, values of $r(\theta_u)$ remain around a similar value for $\theta_u > \theta_u^*$. This would indicate that complexity keeps building up steadily afterwards, and that $\theta_u > \theta_u^*$ (up to $\theta_u = 6\>000$ and perhaps beyond) separate similarly well between simpler and more complex SPEs. 

      We should note, on the one hand, that there are not so many SPEs with that many inhabitants above $\theta^*$. Hence, the complexity spectrum around those values is expected to be less well sampled, and to vary in a stepped fashion, remaining constant for ranges of $\theta_u$ (as seen in Fig.\ \ref{fig:2}{\bf b2}). On the other hand, $\theta^*$ is close to $5\>000$ inhabitants, the threshold chosen in \cite{MussaMira2019}. That choice was guided by Spanish law, which mandates increased urban services in counties (not SPEs) above that population size. This nuance prevents us from suggesting that the law-mandated increase in urban structures might be the cause of the steep complexity buildup at $\theta_u^*$.

    \subsection{An additional build-up of complexity around Dunbar's number}
      \label{sec:res.2}

      The complexity spectrogram in Fig.\ \ref{fig:2}{\bf b2} presents numerous local minima. Each one suggests a population size at which separation between simpler and more complex social networks would be more prominent than in its neighborhood---if our hypotheses hold true. Most of these local minima do not stand out prominently, and might be due to the stochasticity of our system. But at least one such feature sticks out at around $200$ inhabitants, where a smooth, well marked dip seems to form. 

      We locate the corresponding local minimum at $\theta_u^{\dag} \equiv 194$. Fig.\ \ref{fig:2}{\bf a1} shows $c_i$ versus $u_i(\theta_u^\dag)$. This plot illustrates the negative correlation between the speed of the dynamics and the urbanity index, but it also suggests that a linear relationship between those quantities might not be the best descriptor. In most complex systems, relevant properties often scale as power laws \cite{West2017}. If the relationship between the dynamic rate and the urbanity index were of the kind: $c \propto u^\beta$, with $\beta < 0$, this would result in a monotonously decreasing relationship such that more complex SPEs would see an exponentially slower dynamic unfolding. Put otherwise, such scaling also fits our hypotheses. 

      Fig. \ref{fig:2}{\bf b3} shows complexity spectrograms based on the correlation, $\hat{r}(\theta_u)$, between the logarithm of the dynamic rates, $\ln(c_i)$, and the logarithm of the urbanity indexes, $\ln\left(u_i(\theta_u)\right)$. We found that these spectrograms achieve more negative correlations and a much more marked dip at around $\theta_u^\dag$ (precisely at $\hat{\theta}_u^\dag \equiv 183$). Actually, in this representation, $\hat{\theta}_u^\dag$ becomes the global minimum across all scales, and reaches a correlation of $\hat{r}(\hat{\theta}_u^\dag) = -0.68$, smaller than the $r(\theta_u^*) = -0.61$ from the linear representation. This suggests, on the one hand, that $c \propto u^\beta$ is a more natural scaling than the linear one when comparing dynamic rates and urbanity indexes. On the other hand, $\hat{\theta}_u^\dag = 183$ appears as a more marked scale at which a higher, more sudden complexity buildup happens. 

      Most other possible local minima from the linear spectrogram are left behind by the global minimum at $\hat{\theta}_u^\dag$. We still find a dip around $\theta_u^*$ (precisely at $\hat{\theta}_u^* \equiv 4\>735$). Indeed, this dip is now much more pronounced, since $\hat{r}(\theta_u)$ changes steeply both below and above $\hat{\theta}_u^*$. This further suggests that there is a good separation between simpler and more complex social structures at around that size. The spectrogram in log representation displays lower values of $\hat{r}$ for most of the range of $\theta_u$ except, interestingly, for $\theta_u > \theta_u^*$ (green curve in Fig.\ \ref{fig:2}{\bf b3}). 

      We speculate that our marked dip at $\hat{\theta}_u^\dag =183$ might reflect an organic emergence of hierarchy that might take place as human groups become larger than Dunbar's number. The underlying hypothesis here is that the cognitive effort that people can devote to their peers is finite \cite{Dunbar1998, Dunbar2020a}. Dunbar's number was introduced as a speculative, soft upper-bound to the number of meaningful relationships that individuals can sustain. Based on these premises, a simple model of resource allocation predicts different organizations of groups of sizes below and above Dunbar's number \cite{TamaritSanchez2018, TamaritCuesta2022}---an empirically validated feature. 

      Back to opinion dynamics on a social network (such as our process of language shift), in groups of sizes below Dunbar's number everybody can keep each-other as a meaningful relationship. Thus, opinion influences would take place within a single clique---which is a simple kind of graph. For groups of sizes above Dunbar's number, the limitation on meaningful relationships would result in social networks that increasingly present holes and emerging mesoscale motifs---i.e.\ more complex interaction graphs---as meaningful relationships become sparser and clustered. Indeed, groups growing beyond Dunbar's number have been reported to transit from horizontal into hierarchical organizations \cite{Dunbar2020b, Hamilton2007}. According to our hypotheses, such increased complexity would result on a slow-down of social dynamic rates at around Dunbar's scale---as we observe.

    \subsection{Refining the spectrogram through red- and blue-shifts of the spectra}
      \label{sec:res.3} 

      \begin{figure*}
        \centering
        \includegraphics[width=\textwidth]{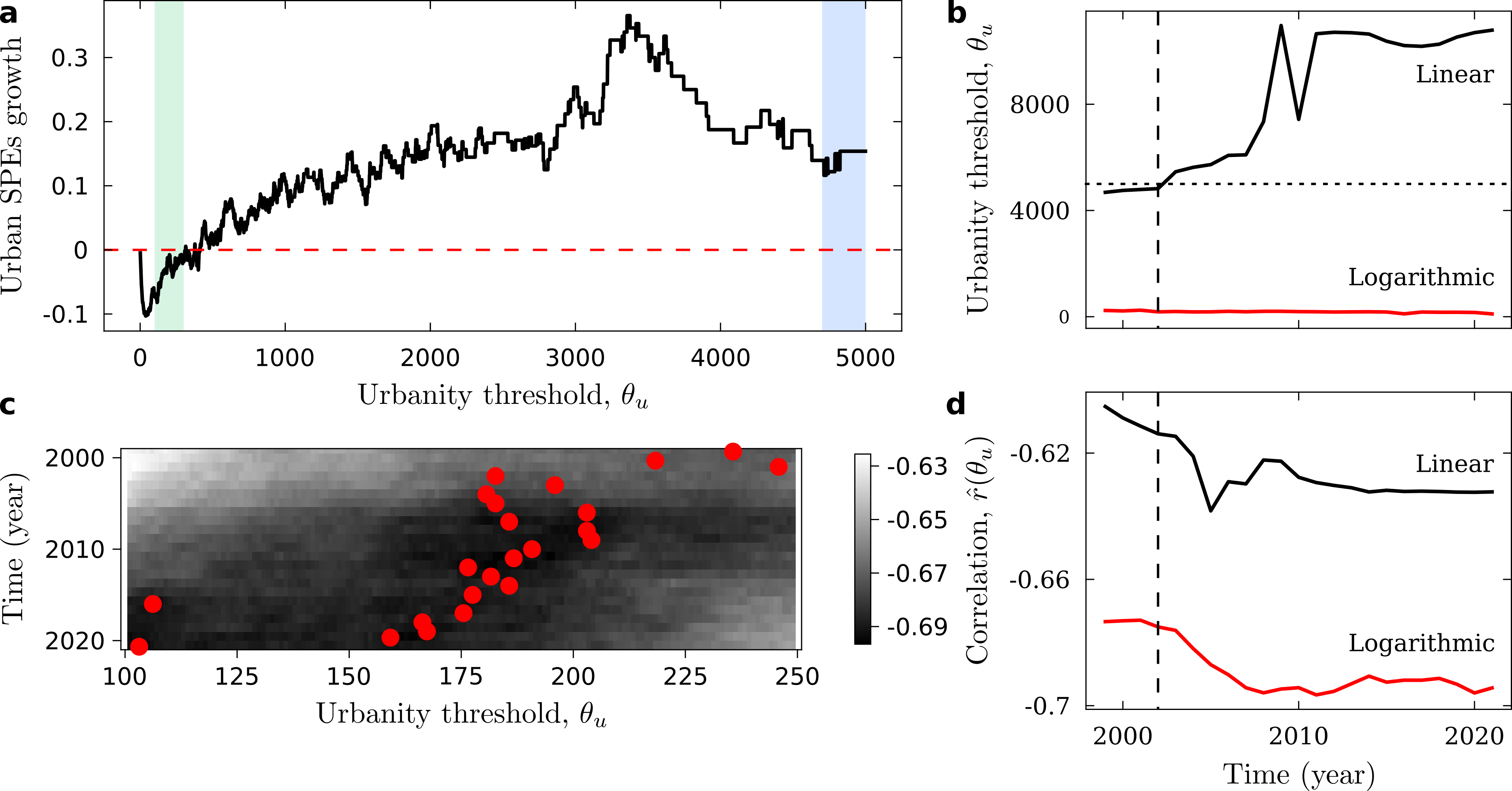}
        \caption{{\bf Red- and blue-shifts in the social complexity spectrum.} {\bf a} Relative growth in the number of SPEs above a given $\theta_u$ between $2002$ and $2011$. This curve reflects how most rural areas have been losing inhabitants during the last decades, while urban areas have been gaining them. These changes in population imply that minima in the spectra will appear shifted with respect to their original position depending on the census year that we choose. Green and blue shading mark respectively the spectral dips at small and large population sizes. {\bf b} Global minima in the linear (black) and logarithmic (red) spectra as a function of census year. According to the gain in population for urban areas, the linear minimum has shifted towards higher values of $\theta_u$. In a simile with light spectroscopy, we call this a blue-shift. {\bf c} A red-shift (towards lower values of $\theta_u$) happens in the logarithmic minimum. This is an average trend---some stochasticity is present in a year-by-year basis. {\bf d} Social groups of a same size might differ in their complexity nevertheless. If simpler groups tend to lose inhabitants (or to gain them more slowly) and more complex ones tend to gain them (or to lose them more slowly), scales separating simpler from more complex social networks would become more outstanding. This seems to be the case in our dataset, as both $r(\theta_u^*)$ and $\hat{r}(\hat{\theta}_u^\dag)$ decrease over time. }
        \label{fig:3}
      \end{figure*} 

      The sociolinguistic dynamics that we study have taken place over most of the XX century. The web of interactions upon which the language shift took place has likely been changing throughout this long period---as people within SPEs died, were born, or migrated. Ideally, we should have a direct measure of each SPE's population during the time-span of our dataset. That would allow us to derive a more sensible, time-composite estimate of the underlying complexity. But unfortunately we cannot retrieve this information. This is likely an important source of noise and inaccuracy in our approach. 

      We trust that our best picture of each SPE's population is given by the 2002 census, published the year after the sociolinguistic polls were collected, which display the social structure of one year before, when the interviews took place. Note that these demographics are a shuffled version of the actual numbers while our language shift dynamics took place. This means that SPEs that were simpler during the studied dynamics might have gained population throughout the XX century, and others that were more complex might have lost population. These movements would result in miss-categorized SPEs that would lower our $r(\theta_u)$ and $\hat{r}(\theta_u)$. 

      If, instead of disarranging the complexity-wise ordering, a series of SPEs would all have gained a similar amount of population, instead of lower $r(\theta_u)$ and $\hat{r}(\theta_u)$ the effect on our spectrogram would be a shift of $\theta_u^{*,\dag}$ and $\hat{\theta}_u^{*,\dag}$ towards higher values. A similar, coordinated loss of inhabitants would result in a shift of $\theta_u^{*,\dag}$ and $\hat{\theta}_u^{*,\dag}$ to lower values with respect to their (unknown) actual positions. Thus, similarly to how spectral bands appear displaced in the light from stars far away, our $\theta_u^{*,\dag}$ and $\hat{\theta}_u^{*,\dag}$ give us an idea of relevant scales. Unlike in the case of distant stars, at the moment we cannot estimate how much our spectra have shifted. Insisting on the metaphor with optimal spectroscopy, we will speak of red-shifts when $\theta_u^{*,\dag}$ or $\hat{\theta}_u^{*,\dag}$ move towards smaller values, and of blue-shifts when they move towards larger ones. 

      There is one such shift that we can study. Note that changes in the SPEs demographics after 2002 do not affect the sociolinguistic dynamics---they have already happened, and are thus somehow fixed. In the last decades, rural Galician areas have lost population, while most urban ones have grown. Fig.\ \ref{fig:3}{\bf a} shows the relative change in the number of {\em urban} SPEs as a function of $\theta_u$ between $2002$ and $2011$. A negative value in this plot means that more SPEs have fallen below a given $\theta_u$, while a positive value means that more SPEs that were smaller than $\theta_u$ in $2002$ had became larger than $\theta_u$ by $2011$. The plot shows the mentioned tendency of very rural areas to lose population, and of urban areas to gain it. The curve is negative, but very close to $0$, at around $\hat{\theta}_u^{\dag}$; and very positive at around $\theta_u^*$. Thus, we would expect a red-shift of the spectral dip at $\hat{\theta}_u^{\dag}$, and a blue-shift of the dip at $\theta_u^*$. 

      Fig.\ \ref{fig:3}{\bf b} and {\bf c} show that this is the case. Fig.\ \ref{fig:3}{\bf c} zooms into the displacement of $\hat{\theta}_u^\dag$ to reveal that, while the red-shift is an average trend over the years, the process is not completely smooth. The stochastic reality of our social system introduces some jittering. 

      Fig.\ \ref{fig:3}{\bf d} shows that both $r(\theta_u^*)$ and $\hat{r}(\hat{\theta}_u^\dag)$ change, actually improving (becoming more negative) in average, as time passes. Focusing on $\hat{\theta}_u^\dag$, this would mean that, among SPEs of a similar size, those that were less complex have tended to lose population below the threshold in that period, while more complex SPEs tended to grow in size---thus $\hat{\theta}_u^\dag$ becoming a better separator of dynamics in simpler versus more complex networks. The best $\hat{r}(\hat{\theta}_u^\dag)$ happens in $2011$ ($\hat{\theta}_u^\dag = 187$, $\hat{r}(\hat{\theta}_u^\dag) = -0.70$). 

      The spectral blue-shift of $\theta_u^*$ is likely only observed until $2007$. After that time, the huge leap in $\theta_u^*$ (consistently defined as the global minimum of the linear spectrum) probably results from that global minimum actually {\em jumping} (not shifting) to other location.

  \section{Discussion}
    \label{sec:disc}

    In this paper we have put forward the {\em social complexity spectrum}, a computational tool to study how certain social dynamics change as a function of the size of the interaction network within which they happen. We argue that, if two hypotheses hold true, our tool allows us to study how social complexity changes as the underlying network of interactions grows---both whether this change is gradual or if it happens in more sudden buildups. One of our hypotheses, inspired by computational results of opinion dynamics on graphs \cite{ToivonenSan2009}, is that the sociolinguistic dynamics of Galician and Castilian Spanish coexistence proceed faster when the web of interactions between speakers is simpler. Our second hypothesis is that population centers with more inhabitants foster more complex social networks. All our empirical results are consistent with what we might expect if these hypotheses hold true. 

    Assuming that our premises are correct, our method quantifies how larger population centers are, indeed, in average, more complex than smaller ones. This is indicated by sustained (at any scale, $\theta_u$) negative correlations, $r(\theta_u)<0$ and $\hat{r}(\theta_u)<0$, between the speed of the sociolinguistic dynamics and our measure of urbanity. Additionally, marked dips of $r(\theta_u)$ and $\hat{r}(\theta_u)$ (i.e.\ singular values $\theta_u^{*,\dag}$ and $\hat{\theta}_u^{*,\dag}$ at which correlation becomes saliently more negative) suggest prominent population sizes at which social complexity builds up more rapidly. 

    One of these singular scales is observed when considering a linear relationship between dynamic rates and urbanity, $c \propto \alpha u$ (with $\alpha<0$ a regression coefficient). This outstanding scale falls near the threshold of $5\>000$ inhabitants chosen by an earlier study \cite{MussaMira2019}. That choice was guided by Spanish law, which mandates an increase of urban equipment in counties larger than $5\>000$ inhabitants. In the future, we hope to perform similar analyses on other countries to validate whether this outstanding scale persists or not, and whether it might correlate with politically-induced changes elsewhere. 

    Another singular scale appears when trying a power-law relationship, $c \propto u^\beta$ (with $\beta<0$ another regression coefficient). This prominent scale ($\hat{\theta}_u^\dag = 183$ inhabitants) is the most salient one throughout all our spectrograms. This suggests, on the one hand, that $\hat{\theta}_u^\dag$ is the threshold that better divides simpler from more complex social networks---or, alternatively, the point at which a steeper buildup of complexity happens. On the other hand, the more negative correlations found for $\hat{r}(\theta_u)$ for most of the $\theta_u$ range ($\hat{r}(\theta_u) < r(\theta_u)$ for $\theta_u < \theta_u^*$) suggests that the power-law scaling $c \propto u^\beta$ is more natural. If this is confirmed, the speed of sociolinguistic dynamics would join many other quantities that scale as a power of population size in urban centers---such as patents, wealth, crime, transmissible diseases, etc.\ \cite{West2017, BettencourtWest2007, BettencourtWest2010, LoboWest2013}. 

    We speculate that $\hat{\theta}_u^\dag$ corresponds to an organic build-up of social complexity as communities cross a size threshold. We suggest that this size threshold is associated to Dunbar's number, a soft limit proposed for the maximal amount of meaningful relationships that humans (and other animals) can sustain \cite{Dunbar1992, Dunbar1998, Dunbar2015, Carron2016,TamaritSanchez2018, Dunbar2020a, BzdokDunbar2022, TamaritCuesta2022}). We argue that this cognitive limit to relationships imposes a constrain on growing social networks that forces them to become more complex---presenting holes, mesoscale communities dictated by affinity or proximity, etc. If this is correct, validating our method with similar dynamics in other regions should see $\hat{\theta}_u^\dag$ invariant. 

    Testing our complexity spectrum in other cases should be a priority. The logical follow-up is to look at similar sociolinguistic dynamics. However, finding good data is relatively difficult. In Galicia, we have the additional advantage that relatively small population centers are abundant and well sampled. We propose the our hypotheses might hold for other kinds of dynamics that require interactions across an extended social group before settling down into their stable configuration. Political shifts and the adoption, and later fade-out, of cultural trends are good candidates. The pervasiveness of virtual social networks affords a unique opportunity to deploy our complexity spectrum---while in such case the speed of social media might interfere with the rate of the dynamics themselves. 

    Another sense in which we hope to extend this methodology is towards larger population sizes: Are additional complexity increases gradual? Or are there new singular scales at which sudden buildups are observed? The size of the Galician population can only test our tools up to some dozen thousand inhabitants, while cities around the world well surpass the dozen million mark. Finding singular scales of social complexity buildup is relevant to understand, manage, and engineer more optimal societies. Steep changes in the underlying social structure inform us of qualitative elements that might be needed when modeling such systems. For example, if our speculation is correct, smaller groups could be modeled as cliques, while a rich mesoscale with holes is needed already for moderately large communities. The scale $\hat{\theta}_u^\dag$ would inform us of when we need to change our modeling approach. 

    An interesting possibility arises if we relax our second hypothesis. We assume that larger population centers result in more complex interaction networks. What if this is not the case? An ingredient for internal complexity is certain sustained degree of heterogeneity. It is possible to imagine a city large enough, and managed from an homogenizing perspective, such that the diversity of its constituent communities is erased. If this were the case, we speculate that such larger cities would have {\em less} complex interaction networks than smaller ones. Then, if our other hypothesis holds (i.e.\ that certain dynamics run faster over simpler interaction graphs), we would expect a positive correlation when testing the relationships between dynamic rates and urbanity. That would be reflected as peaks in our spectrogram. Their existence would point towards limits (perhaps of a physical or computational nature) to the growing complexity of social networks. This possibility connects ultimately with the phenomenon of {\em complexity drain} \cite{McShea2002}, by which certain systems present a conflict between their complexity and that of their parts. For example, species of ants with relatively small nests have more complex individual ants while their emergent repertoire is more limited. On the other hand, species that form larger communities present a much higher versatility at the super-organism level, while individuals are much simpler (i.e.\ more specialized and with limited capacities) \cite{OsterWilson1978}. 

    A similar inversion (from dips to peaks) in our spectra would be observed if we could prove social processes that happen faster in more complex social networks. Speculatively, explosive percolation \cite{AchlioptasSpencer2009} comes to mind. This is an extremely fast kind of phase transition, that is known to be induced when attempts are made to slow it down \cite{DSouzaNagler2015}. In our context, slower social dynamics in more complex communities could play the role of such deliberate efforts to slow down change, but eventually prompt a much faster (indeed explosive) shift. Abusing the analogy, revolutionary processes are often kick-started in urban settings. 

    A final, important test for our ideas is to correlate these results with direct studies of social structure---which might not be an easy enterprise. Because of this, we think that our spectrum offers a great chance to study phenomena around social complexity and how it scales with the size of a community. 

\vspace{0.2 cm}

  \section*{Acknowledgments}

    The authors wish to acknowledge Juan Jos\'e V\'azquez-Portome\~ne Seijas, state counsel of Spain, for his assistance to understand Spanish law properly. We also wish to thank Jose Cuesta, from the Carlos III University of Madrid, for useful comments on an earlier draft of this article.

\vspace{0.2 cm}

  \section*{Funding}

    LFS received funding from the Spanish National Research Council (CSIC) and the Spanish Department for Science and Innovation (MICINN) through a Juan de la Cierva Fellowship (IJC2018-036694-I) and grant PID2020-113284GB-C21, funded by MCIN/AEI/10.13039/501100011033. The Spanish MICINN has also funded the ``Severo Ochoa'' Centers of Excellence distinction (grant SEV 2017-0712) for the CNB, where LFS carried out his research. APM received funding from the Spanish Ministerio de Economía y Competitividad and European Regional Development Fund under contract RTI2018-097063-B-I00 AEI/FEDER, UE, and from Xunta de Galicia under Research Grant No. 2021-PG036. All these programs are co-funded by FEDER (UE). The simulations were run in the Supercomputer Center of Galicia (CESGA) and we acknowledge their support.

\end{document}